\documentclass[preprint2]{emulateapj}

\def\etal{{et~al.}}

\def\sec{\hbox{${}^{\prime\prime}$}}
\def\lae{\mathrel{<\kern-1.0em\lower0.9ex\hbox{$\sim$}}}
\def\gae{\mathrel{>\kern-1.0em\lower0.9ex\hbox{$\sim$}}}
\def\msun{${\cal M}_\odot$}
\def\etal{et~al.~}

\def\kms{km~s$^{-1}$}

\def\msb{${\cal M}_{\rm SBH}$-$\sigma$~}

\def\mmc{${\cal M}_{\rm CMO}$-${\cal M}_{\rm gal}$~}

\shorttitle{A Relation Between Stellar Nuclei and SBHs}

\shortauthors{Ferrarese \etal}

\begin{document}

\title{A Fundamental Relation Between Compact Stellar Nuclei,
  Supermassive Black Holes, and Their Host Galaxies\altaffilmark{1}}

\author{Laura Ferrarese\altaffilmark{2},
Patrick C\^ot\'e\altaffilmark{2,3},
Elena Dalla Bont\`a\altaffilmark{2,4},
Eric W. Peng\altaffilmark{2,3},
David Merritt\altaffilmark{5},
Andr\'es Jord\'an\altaffilmark{6,7},
John P. Blakeslee\altaffilmark{8},
Monica Ha\c{s}egan\altaffilmark{3,9},
Simona Mei\altaffilmark{10},
Slawomir Piatek\altaffilmark{11},
John L. Tonry\altaffilmark{12},
Michael J. West\altaffilmark{13}}

\altaffiltext{1}{Based on observations with the NASA/ESA {\it Hubble
Space Telescope} obtained at the Space Telescope Science Institute,
which is operated by the Association of Universities for Research in
Astronomy, Inc., under NASA contract NAS 5-26555.}
\altaffiltext{2}{Herzberg Institute of Astrophysics, National
Research Council of Canada, 5071 West Saanich Road, Victoria, BC,
V8X 4M6, Canada; laura.ferrarese@nrc-cnrc.gc.ca}
\altaffiltext{3}{Visiting Astronomer, KPNO/NOAO, which is operated by AURA under
cooperative agreement with the NSF.}
\altaffiltext{4}{Dipartimento di Astronomia, Universit\'a di Padova, 
Vicolo dell'Osservatorio 2, 35122 Padova, Italy}
\altaffiltext{5}{Dept. of Physics, Rochester Institute of
Technology, 84 Lomb Memorial Drive, Rochester, NY 14623}
\altaffiltext{6}{European Southern Observatory, Karl-Schwarzschild-Str.
2, 85748 Garching, Germany}
\altaffiltext{7}{Astrophysics, Denys Wilkinson Building, University
of Oxford, 1 Keble Road, Oxford, OX1 3RH, UK}
\altaffiltext{8}{Dept. of Physics, Washington State University,
Webster Hall 1245, Pullman, WA 99164-2814}
\altaffiltext{9}{Dept. of Physics and Astronomy, Rutgers
University, New Brunswick, NJ 08854}
\altaffiltext{10}{Dept. of Physics and Astronomy, The Johns
Hopkins University, 3400 North Charles Street, Baltimore, MD
21218-2686}
\altaffiltext{11}{Dept. of Physics, New Jersey Institute of Technology,
Newark, NJ 07102}
\altaffiltext{12}{Institute for Astronomy, University of Hawaii, 2680
Woodlawn Drive, Honolulu, HI 96822}
\altaffiltext{13}{Dept. of Physics and Astronomy, University of
Hawaii, Hilo, HI 96720}

\slugcomment{Accepted by the {\it Astrophysical Journal Letters.}}

\begin{abstract}
Imaging surveys with the {\it Hubble Space Telescope} ({\it HST}) have
shown that $\approx$ 50--80\% of low- and intermediate-luminosity
galaxies contain a compact stellar nucleus at their center, regardless
of host galaxy morphological type. We combine {\it HST} imaging for
early-type galaxies from the {\it ACS Virgo Cluster Survey} with
ground-based long-slit spectra from KPNO to show that the masses of
compact stellar nuclei in  Virgo Cluster galaxies obey a tight
correlation with the masses of the host galaxies. The same correlation
is obeyed by the supermassive black holes (SBHs) found in
predominantly massive galaxies.  The compact stellar nuclei in the
Local Group galaxies M33 and NGC 205  are also found to fall along
this same scaling relation.  These results indicate that a generic
by-product of galaxy formation is the creation of a {\it central
massive object} (CMO) --- either a SBH or a compact stellar nucleus
--- that contains a mean fraction, $\approx$ 0.2\%, of the total
galactic mass. In galaxies with masses greater than ${\cal M_{\rm
gal}} \sim$ a few $10^{10}{\cal M}_{\odot}$,\ SBHs appear to be the
dominant mode of CMO formation.
\end{abstract}

\keywords{black hole physics--galaxies: elliptical and
lenticular--galaxies: nuclei--galaxies: structure --galaxies:
kinematics and dynamics}

\section{Introduction}
\label{sec:introduction}

Stellar and gas dynamical studies in an ever-increasing number of
galaxies have established that many --- and perhaps all --- luminous
galaxies contain central supermassive black holes (SBHs). Following
the discovery that the SBH masses, $\cal M_{\rm SBH}$, correlate with
various properties of the host galaxy --- such as bulge luminosity
(Kormendy \& Richstone 1995),  mass (H{\"a}ring \& Rix 2004), velocity
dispersion (Ferrarese \& Merritt 2000; Gebhardt \etal\ 2000a), light
concentration  (Graham \etal\ 2001),  and halo circular velocity
(Ferrarese 2002)  --- it has become widely accepted that SBH and
galaxy formation are closely entwined.

Unfortunately, the physical mechanisms underlying this connection
remain obscure (e.g. Silk \& Rees 1998; Portegies Zwart \etal\ 2004;
Shapiro 2005).  Despite intense observational effort, only about 30
galaxies have secure SBH detections (see the recent review of
Ferrarese \&  Ford 2005), the great majority of which are luminous
galaxies with magnitudes in the range $-22 \lesssim M_B \lesssim -18$.
It is unclear if fainter and less massive galaxies {\it also}
contain SBHs and, if so, whether such objects would obey
extrapolations of the SBH scaling relations defined by the bright
galaxies. Searches for SBHs in  low-luminosity members of the Local
Group have so far produced ambiguous results. There is no evidence for
a SBH in either M33 (Merritt \etal\ 2001; Gebhardt \etal\ 2001) or NGC
205 (Valluri \etal\ 2003), yet M32 does appear to contain a SBH with
$\cal M_{\rm SBH}$ $\approx$ 2.5$\times10^6{\cal M}_{\odot}$
(Verolme \etal\ 2002).

Although they have very different morphologies, M33 (Sc II-III) and
NGC 205 (S0/E5pec) share one noteworthy similarity: their centers are
both marked by the presence of a compact stellar nucleus (with
half-light radius $r_h \lesssim$ 2-4~pc) that is $\sim$~20 times
brighter than a typical globular cluster (e.g.,  Kormendy \& McClure
1993; Butler \& Martinez-Delgado 2005). While ground-based surveys of
the Virgo and Fornax Clusters had shown $\sim$ 25\% of dE galaxies to
contain such nuclei (e.g.  Binggeli, Tammann \& Sandage 1985; Ferguson
1989; Binggeli \& Cameron 1991), recent observations with the {\it
Hubble Space Telescope} ({\it HST})  have revealed them to be far more
common. About 50-70\% of late-type galaxies observed by {\it HST}
contain a distinct nuclear star cluster (Carollo, Stiavelli \& Mack
1998; Matthews et~al 1999; B\"oker et~al. 2002, 2004; Balcells \etal\
2003), while a recent {\it HST} survey of 100 galaxies in the Virgo
Cluster has detected nuclei in a comparable fraction (66--82\%) of
early-type galaxies (C\^ot\'e \etal\ 2006; see also Lotz \etal\ 2004;
Graham \& Guzman 2003; Grant \etal\ 2005).

In this {\it Letter}, we explore the connection between compact
stellar nuclei, SBHs and their host galaxies by combining {\it HST}
imaging for 100 early-type galaxies from the {\it ACS Virgo Cluster
Survey} (ACSVCS; C\^ot\'e \etal\ 2004) with new ground-based long-slit
spectra for the brightest 69 of these galaxies.  We show that the mass
of the Central Massive Object (CMO) --- either a compact stellar
nucleus or a SBH --- scales in direct proportion to the galaxy mass.
This finding
points to a direct link between SBHs, which are preferentially
detected in the brightest galaxies, and the compact stellar nuclei
commonly observed in galaxies of low and intermediate luminosity.

\section{Observations and Data Reductions}
\label{sec:data}

{\it HST} images for 100 members of the Virgo Cluster were
acquired with the {\it Advanced Camera for Surveys} ({\it ACS}; Ford
\etal\ 1998) as part of the ACSVCS (GO-9401).  The program galaxies
span a range of $\approx$ 460 in blue luminosity and have early-type
morphologies: E, S0, dE, dE,N or dS0.  Images were taken in WFC mode
with a filter combination roughly equivalent to the $g$ and $z$ bands
in the SDSS photometric system.  The images cover
a $\approx$ 200$^{\prime\prime}\times200^{\prime\prime}$ field with
$\approx 0\farcs1$ resolution and 0\farcs05 pixel$^{-1}$ sampling.
For each galaxy, azimuthally averaged surface brightness profiles were
determined as described  by Ferrarese \etal\ (2006) and C\^ot\'e
\etal\ (2006). We refer the reader to these papers for full details
of the analysis.

\begin{figure}
\plotone{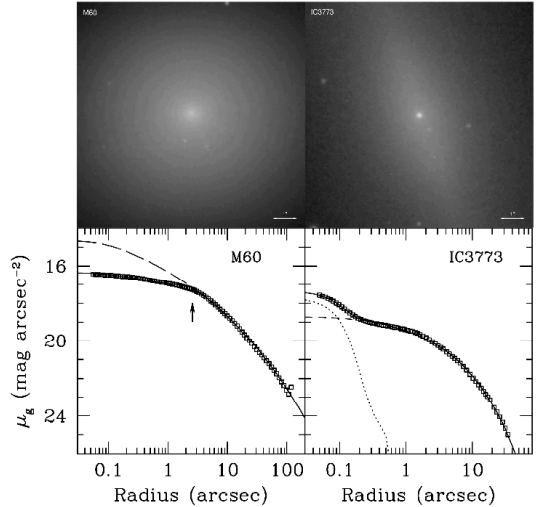}
\caption{ {\it (Upper Panels)} $g$-band images showing the central
regions of M60 (left) and IC 3773 (right), the 3rd and 51st brightest
galaxies respectively in the ACSVCS.   {\it (Lower Panels)}
Azimuthally-averaged $g$-band surface brightness profiles for the same
two galaxies. M60 is a typical non-nucleated ``core-S\'{e}rsic"
galaxy: the best core-S\'{e}rsic model is shown as a solid curve. The
vertical arrow shows the radius, $r_b$, at which the outer S\'{e}rsic
profile ``breaks" to an inner power-law; the long-dashed curve shows
the inward extrapolation of the S\'{e}rsic model fitted to the data
beyond $r_b$. For IC 3773, we show the best-fit model which consists
of a central King model for the nucleus (dotted curve) and a
S\'{e}rsic model for the underlying galaxy (dashed curve). The solid
curve shows the composite model.
\label{fig:sbp}}
\end{figure}

The 11 ACSVCS galaxies brighter than $M_B \approx -20$ mag are found
to have surface brightness profiles that are accurately represented by
a ``core-S\'{e}rsic" model (Graham \etal\ 2003; Trujillo \etal\ 2004),
described by a S\'ersic  (1968) model outside a ``break radius",
$r_b$, of a few arcseconds, and a shallower power-law interior to
$r_b$. None of these bright galaxies shows clear evidence for a
central stellar luminosity excess over the fitted profile. In
contrast,  nearly all of the fainter galaxies are well fitted with
pure S\'{e}rsic models; in addition, 60--80\% of these 89 galaxies
show evidence for a nucleus, identified as a luminosity excess over
the best fitted profile within $\sim$~1\arcsec. In 51 galaxies, the
nucleus is conspicuous enough to allow us to measure photometric and
structural parameters; we do so by adding a King model (King 1966) to
the S\'ersic component when fitting the surface brightness profile.
Fig.~\ref{fig:sbp} shows images and surface brightness profiles for
two representative galaxies from the ACSVCS: a ``core-S\'{e}rsic"
galaxy with $M_B \approx -21.4$ mag (M60), which also happens to have
a dynamically measured SBH mass  ${\cal M}_{\rm SBH}$ =
2.0($^{+0.5}_{-0.6}$)$\times10^8{\cal M}_{\odot}$ (Gebhardt \etal\
2003); and a typical nucleated S\'{e}rsic galaxy  (IC 3773) with $M_B
\approx -17.3$ mag.

Although compact, the central nuclei are resolved in all but a half
dozen or so cases: half-light radii range from $r_h \le 2$ pc (i.e.,
unresolved) to 62 pc, with a median of $\approx$ 4~pc.  We estimate
total masses for the nuclei by multiplying their $g,z$ luminosities
(determined by integrating the best-fit King models) with appropriate
mass-to-light ratios, $\Upsilon_g$ and $\Upsilon_z$.  Single-burst
stellar population models from Bruzual \& Charlot (2003) were used to
estimate $\Upsilon_g$ and $\Upsilon_z$ for each nucleus, at the
metallicity appropriate for the observed color, for a fixed assumed
age of $\tau$ = 5~Gyr and adopting a Chabrier (2003) IMF. The
uncertain ages of the nuclei is the dominant source of uncertainty on
the derived masses; the difference (of order $\approx\pm$45\%) between
the 5~Gyr masses and those obtained assuming ages of 2 and 10~Gyr, is
taken as representative of the error on the quoted values.

Long-slit spectra for the 69 ACSVCS galaxies brighter than $M_B =
-16.5$ mag, of which 29 are classified as certainly nucleated, were
obtained  between 2003 March 10--12 and 2003 March 21--28 using
facilities at the Kitt Peak National Observatory (KPNO). All spectra
were obtained  with the slit oriented along the galaxy photometric
major axis, and were centered on the Mg b triplet near~5200~\AA. Three
separate instrumental setups were used for th`<e bright, middle and
faint thirds of the sample. Spectral resolutions ranged between 94
\kms~ and 220~\kms~at 5200~\AA. Exposure times ranged between 2400 s
and 5400 s.  Between one and three giant or subgiant stars of spectral
type G8--K2, to be used as velocity dispersion templates, were
observed each night  with the same instrumental setup adopted for the
galaxies.

Systemic velocities, $v$, and velocity dispersions, $\sigma$, were
extracted  using the Penalized Pixel-Fitting code of Cappellari \&
Emsellem (2004) from spectra binned, in the spatial direction,  within
an aperture of radius equal to the galaxy effective radius, $R_e$.
The final $v$ and $\sigma$, and their errors, are the averages and
standard deviation of the values obtained using three different
template stars.

\section{Results}
\label{sec:results}

\begin{figure*}
\plotone{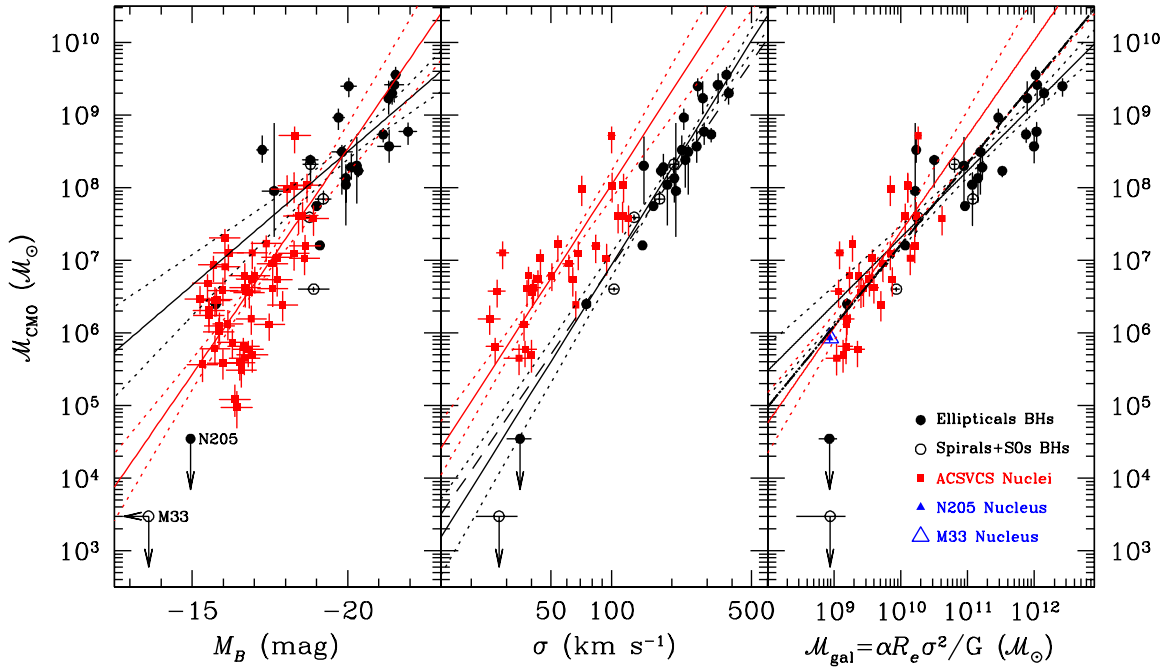}
\caption{ {\it (Left Panel)} Mass of the Central Massive Object (CMO)
plotted against absolute blue magnitude of the host galaxy (or bulge
for spiral galaxies). Nuclei from the ACSVCS are shown as red
squares. The supermassive black holes (SBHs) in early-type  and spiral
galaxies are  shown as filled and open circles respectively.   {\it (Middle
Panel)} CMO mass as a function of velocity dispersion of the  host
galaxy, measured within $R_e$.  {\it (Right Panel)} CMO mass plotted
against galaxy mass, defined as ${\cal M}_{\rm gal} \equiv
{\alpha}R_e\sigma^2/G$ with $\alpha = 5$. In
all panels, the solid red and black lines show the best fits to the 
nuclei and early-type SBH samples respectively,
with 1$\sigma$ confidence levels shown by the dotted
lines. In the middle panel, the dashed line is the best fit
\msb~relation of Tremaine et al.\ (2002).  In the right panel, the
dashed line is the fit obtained for the combined nuclei+SBH
sample. Coefficients for all fits are listed in Table~1.
\label{fig:ms}}
\end{figure*}

In the left panel of Fig.~\ref{fig:ms} masses for the nuclei are shown
in red, plotted as a function of the extinction-corrected, absolute
blue magnitude of the galaxy $M_B$ (Binggeli  et al.\ 1985). In the
middle panel, masses are plotted against the stellar velocity
dispersion $\sigma$, measured within $R_e$.  Solid black dots show
SBH masses (from Table~II of Ferrarese \& Ford 2005; $M_B$ values are
mostly from the RC3, de Vaucouleurs et al.\ 1991) detected based on
stellar/gas dynamical studies which resolve the sphere of influence.
Velocity dispersions for the galaxies with SBHs are from Tremaine et
al.\ (2002) or (for the three galaxies for which no values were
published in Tremaine et al.) Ferrarese \& Ford (2005).

A first noteworthy point is that there is almost no overlap in the
range of $M_B$ and $\sigma$ occupied by galaxies with nuclei and SBHs,
with the former found preferentially in fainter galaxies with lower
velocity dispersion. This is partly, but not entirely, due to
observational biases. Galaxies brighter than  $M_B \sim -20$ mag do
not contain nuclei (see \S\ref{sec:data}) but host
SBHs\footnotemark. Evidence supporting the latter statement comes from
the fact that SBHs have been successfully detected in all galaxies
targeted in this magnitude range by dynamical studies, and  masses
have been found to be consistent with the \msb~relation.  Furthermore,
the existence of SBHs in all bright galaxies is required to reconcile
the local and AGN SBH mass functions (e.g. Marconi \etal\ 2004;
Shankar \etal\ 2004).

\footnotetext{The central light excess in the bright core-S\'ersic
galaxies  M87 (e.g. Ferrarese et al.\ 2006)  and NGC 6166 (Capetti et
al.\ 2000),  both of which are strong radio sources, are unresolved
and have a non-stellar origin.}

As one moves to fainter galaxies, nuclei become increasingly common,
and are almost always present in galaxies fainter than $M_B \sim -18$
mag (C\^ot\'e et al.\ 2006).  Located at the faint end of the
magnitude range spanned by the ACSVCS galaxies, NGC 205 and M33, both
of which are strongly nucleated, are not believed to contain SBHs
(Merritt \etal\ 2001; Gebhardt \etal\ 2001; Valluri \etal\
2003). Galaxies with nuclei, SBHs or, possibly, both, 
are found in the $-18 \lesssim M_B
\lesssim -20$ mag range. In the same magnitude range, galaxies exist
for which the existence of a SBH is uncertain (e.g. NGC 3379, NGC
4342, Gebhardt \etal\ 2000b; Cretton and van den Bosch 1999).  
Overall, therefore, the existing data supports
a view in which {\it bright galaxies often, and perhaps always,
contain SBHs but not stellar nuclei. As one moves to fainter galaxies,
nuclei become the dominant feature while SBHs might become less
common, and perhaps disappear entirely at the faint end.}

Further insight can be gained by relating the masses of nuclei and
SBHs to the properties of the host galaxy. Regression fits  (Akritas
\& Bershady 1996) for the early-type galaxies (spirals have been
excluded to ensure consistency with the ACSVCS sample) confirm the
visual impression from the left and middle panels of Fig~\ref{fig:ms}
that nuclei and SBHs obey statistically different scaling relations
with respect to both galaxy magnitude and bulge velocity dispersion
(Table 1).  However, a different picture emerges (see the right panel
of  Fig~\ref{fig:ms}) when the virial mass of the host galaxy,
${\cal M}_{\rm gal} = \alpha R_e \sigma^2/G,$ is considered. Here
$G$ is the gravitational constant and $\alpha = 5$ (Cappellari et al.\
2006); the geometric effective radius, $R_e = a\sqrt{1-\epsilon}$, $a$
being the radius measured along the isophotal semi-major axis, is
taken from Ferrarese \etal\ (2006) for the ACSVCS galaxies, and from
Marconi \& Hunt (2003) for the galaxies hosting SBHs.  A regression
analysis (Table~1) demonstrates that {\it  the nuclei and SBHs obey a
common scaling relation linking their mass to the virial mass of the
host galaxy\footnotemark}. Furthermore, the same relation is obeyed by
the nuclei of NGC 205 (Geha \etal\ 2006)  and M33 (for which we adopt
$R_e = 1$ kpc, intermediate to the values of Minniti et al.\ 1994 and
Regan \& Vogel 1994), while the upper limits on the mass of the
central SBH in both galaxies fall well below the best fit line.

\footnotetext{We note that all nucleated galaxies are early-type
(ellipticals and S0s), and so are  the galaxies with central SBHs used
in performing the fits.  For all, ${\cal M}_{\rm gal}$ is a measure of
the total mass of the galaxy, rather than the mass of the bulge, since
even in lenticular galaxies, both bulge  and disk components
contribute to the measured $\sigma$.}

On these grounds,  we suggest that SBHs and nuclei should be grouped
together under the terminology, ``Central Massive Object'' (CMO),
which we adopt for the remainder of this {\it Letter}.  Constraining
the slope of the \mmc~relation to be unity leads to a constant ratio
between CMO and galaxy mass ${\cal M}_{\rm CMO}/{\cal M}_{\rm gal}
\approx 0.18\%$ (with a $\pm1\sigma$  range of 0.06--0.52\%), a
conclusion also reached, based on photometric data only, by C\^ot\'e
et al. (2006) and Wehner \& Harris (2006).  We note that our
conclusions are insensitive to the exact methodology used in measuring
$\sigma$.  In particular, integrating $\sigma$  within 1\sec~(a region
dominated by the nucleus), between 3\sec~and $R_e$ (a region dominated
by the host galaxy), or within 1/8$R_e$ (as in Ferrarese \& Merritt
2000), changes the individual measurements (by, on average  $\sim$5\%,
although differences of up to 30\% can be seen for some of the fainter
galaxies) but does not alter the overall trend.

\section{Discussion}
\label{sec:discussion}

\tabletypesize{\tiny}
\begin{deluxetable}{lcccc}
\tablecaption{Scaling Relations for Central Massive Objects (CMOs)}
\tablecolumns{5}
\tablehead{
\colhead{($X,Y$)} &
\colhead{$a$} &
\colhead{$b$} &
\colhead{$\chi^2_r$} &
\colhead{$N$} 
}
\startdata
$M_B+19.9$ mag,${\cal M}_{\rm SBH}$                       &           $-$0.37 $\pm$ 0.08 & 8.46 $\pm$ 0.11  &  19.1  & 21\\
$M_B+16.9$ mag,${\cal M}_{\rm nuc}$                       &           $-$0.62 $\pm$ 0.10 & 6.59 $\pm$ 0.09  &  \phantom{1}3.1  & 51\\
$\sigma/(224{\rm~km~s^{-1})},{\cal M}_{\rm SBH}$               & \phantom{$-$}4.41 $\pm$ 0.43 & 8.48 $\pm$ 0.07  &   \phantom{1}3.0  & 21\\
$\sigma/(54{\rm~km~s^{-1})},{\cal M}_{\rm nuc}$               & \phantom{$-$}4.27 $\pm$ 0.61 & 6.91 $\pm$ 0.11  &  \phantom{1}7.0  & 29\\
${\cal M}_{\rm gal}/(10^{11.3}$ \msun),${\cal M}_{\rm SBH}$ & \phantom{$-$}0.92 $\pm$ 0.11 & 8.47 $\pm$ 0.08  &   \phantom{1}8.9  & 21\\
${\cal M}_{\rm gal}/(10^{9.6}$ \msun),${\cal M}_{\rm nuc}$ & \phantom{$-$}1.32 $\pm$ 0.25 & 6.91 $\pm$ 0.09  &  \phantom{1}6.0  & 29\\
${\cal M}_{\rm gal}/(10^{10.3}$ \msun),${\cal M}_{\rm CMO}$ & \phantom{$-$}1.12 $\pm$ 0.07 & 7.57 $\pm$ 0.07  &  \phantom{1}8.9  & 50\\
${\cal M}_{\rm gal}/(10^{10.3}$ \msun),${\cal M}_{\rm CMO}$ & $\equiv$ 1                   & 7.56 $\pm$ 0.47  &  \phantom{1}7.4  & 50\\
\enddata
\tablecomments{Columns (2) and (3) give best-fit coefficients for
linear relations of the form log$Y = a$log$X+b$ (or log$Y = aX+b$ 
when fitting to the galaxies' magnitudes). $\chi^2_r$ is the reduced 
$\chi^2$ of the fit, while $N$ is the number of datapoints used in the fit.}
\end{deluxetable}
\tabletypesize{\normalsize}

The main finding in this paper is that a common  \mmc~relation leads
smoothly from SBHs to nuclei as one moves down the mass function for
early-type galaxies.  This suggests that a single mechanism is
responsible for the growth --- and perhaps the formation --- of both
nuclei and SBHs. It also points to galaxy mass as the primary (though
not necessarily only) parameter regulating such growth.

Stellar cusps and SBHs have often been linked in the literature, and
it is therefore natural to ask whether nuclei could be the by-product
of SBH evolution. Since most nuclei in the ACSVCS galaxies are
spatially resolved, we can exclude that they formed either via
adiabatic  growth (Young 1980) or, in the fainter galaxies, via the
Bahcall-Wolf  (1976) process, since either mechanism generates a
power-law cusp only within a fraction of the SBH's influence radius
(Merritt \& Szell 2005).  However, nuclei and SBHs might coexist in
{\it some} galaxies; the most promising cases being M32 (Verolme 
et al. 2002) and the Milky Way 
(Ghez et al. 2003; Sch\"odel et al.\ 2003). Although neither
galaxy possesses the kind of nucleus seen in the  faintest
ACSVCS galaxies, it is quite possible, if not likely, for nuclei to
undergo structural changes as a consequence of the presence of the
central SBH.

Beyond this, the exact interplay between nuclei and SBHs remains
elusive.  It is possible that nuclei form in all galaxies, but in the
most massive systems either they subsequently collapse to SBHs, or are
destroyed or modified by the evolution of pre-existing SBHs. As
mentioned in \S\ref{sec:data}, the surface brightness profiles of
galaxies brighter than  $M_B \sim -20$ mag, which host SBHs but not
nuclei, display an inner ``deficit'' relative to the inward
extrapolation of the S\'ersic law that best fits the outer parts
(Fig.~\ref{fig:sbp};  Ferrarese \etal\ 2006; Trujillo \etal\ 2004).
Such deficits are generally believed to result from the disruptive
effects of binary SBH evolution (Ebisuzaki, Makino \& Okumura 1991;
Milosavljevi\'c \etal\ 2002; Ravinandrath \etal\ 2002; Graham 2004),
so the same process might have led to the destruction of a central
nucleus. Nuclei in slightly fainter galaxies which also contain a SBH
might have avoided destruction or might have been regenerated at a
later time, perhaps, e.g., by subsequent star formation. Determining
stellar  population ages for the nuclei would provide some
observational constraints for this scenario.  Age and abundance
measurements for our nuclei will be presented in a future ACSVCS paper.

Alternatively, the formation of SBHs and nuclei could be mutually
exclusive, with only material collected at the centers of  massive
systems able to collapse to a black hole, while in less massive
galaxies the collapse is halted and a star cluster is formed. SBHs and
nuclei are almost certainly mutually exclusive in the faintest
galaxies considered here, as suggested by the fact that, although the
nuclear masses of NGC 205 and M33 are fully consistent with the
\mmc~relation, the upper limits on their SBH masses are {\it not},
implying that neither galaxy contains a SBH of the sort expected from
extrapolations of the scaling relations defined by SBHs in massive
galaxies. If the formation of a SBH prevents the formation of a ``NGC
205-type'' nucleus (or vice versa), then nuclei of galaxies which are
known to host SBHs (e.g., M32,  the MW, and potentially all galaxies
with a few $\times 10^9$~\msun~$\lesssim {\cal M}_{\rm gal} \lesssim$
a few$\times 10^{10}$~\msun) would necessarily have to belong to a
separate class. A high resolution study of the nuclear morphology in
nearby ($d \lesssim 15$ Mpc) galaxies might unveil whether nuclei in
galaxies of different mass are structurally distinct. These issues
will be explored in more detail in forthcoming papers.

\acknowledgments

We thank the referee, Alister Graham, for many useful comments.
Support for program GO-9401 was provided through a grant from
STScI, which is operated by AURA under NASA contract NAS5-26555.

\end{document}